\documentclass{camera}
\usepackage{graphicx}  
\begin{document}
\newcommand{\goo}{\,\raisebox{-.5ex}{$\stackrel{>}{\scriptstyle\sim}$}\,}
\newcommand{\loo}{\,\raisebox{-.5ex}{$\stackrel{<}{\scriptstyle\sim}$}\,}

%
\title{From nuclear multifragmentation reactions\\ to supernova explosions}

%
\author{Igor N. Mishustin}

%
\organization{FIAS, J.W. Goethe University, 60438 Frankfurt am Main\\and\\
Kurchatov Institute, Russian Research Center, 123182 Moscow}

\maketitle

\begin{abstract}
In this talk I discuss properties of hot stellar matter at
sub-nuclear densities which is formed in supernova explosions. 
I emphasize that thermodynamic conditions there are rather
similar to those created in the laboratory by intermediate-energy
heavy-ion collisions. Theoretical methods developed for the
description of multi-fragment final states in such reactions can
be used also for description of the stellar matter. I present main
steps of the statistical approach to the equation of state and
nuclear composition, dealing with an ensemble of nuclear species
instead of one "average" nucleus. 

\end{abstract}

\section{Introduction}

A type II supernova explosion is one of the most spectacular
events in astrophysics, with huge energy release of about
$10^{53}$ erg or several tens of MeV per nucleon \cite{Bethe}.
When the core of a massive star collapses, it reaches densities
several times larger than the normal nuclear density $\rho_0=0.15$
fm$^{-3}$. The repulsive nucleon-nucleon interaction gives rise to
a bounce-off and formation of a shock wave propagating through the
in-falling stellar material, predominantly Fe.
Hydrodynamical simulations (see e.g. refs.
\cite{Janka,Thielemann}) show that during the collapse and
subsequent explosion the temperatures $T\approx (0.5\div 10)$ MeV
and baryon densities $\rho_B \approx (10^{-5}\div 2) \rho_0$ can
be reached. A schematic view of the post-collapse star core is
presented in Fig.~1. 

\begin{figure}
\includegraphics[width=12cm]{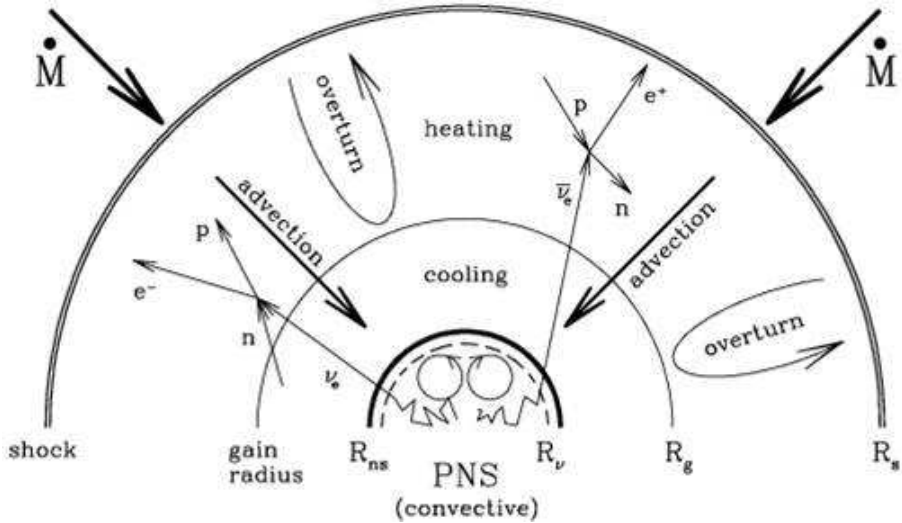}
\caption{\small{Schematic view of the post-collapse stellar core
230 ms after the bounce-off, as predicted by the hydrodynamical
simulations \cite{Janka}. The neutrino heating and convection
processes help to revive the shock. Region between the
protoneutron star (PNS) and the shock front is called the Hot
Bubble. In-falling matter is represented by thick arrows labelled
by $\dot M$.} }
\end{figure}

For the realistic description of supernova physics one should
certainly use experience accumulated in recent years by studying
intermediate-energy nuclear reactions. In particular,
multifragmentation reactions provide valuable information about
hot nuclei in dense environment. According to present
understanding, based on numerous theoretical and experimental
studies of multifragmentation reactions, prior to the break-up a
transient state of nuclear matter is formed, where hot nuclear
fragments exist in equilibrium with free nucleons. This state is
characterized by a certain temperature $T \sim 3-6$  MeV and a
density which is typically 3-5 times smaller than the nuclear
saturation density $\rho_0$. A very good
description of such systems is achieved with the Statistical
Multifragmentation model (SMM), for a review see ref. \cite{SMM}.
The statistical nature of multifragmentation is confirmed by
numerous experimental observations, e. g. "rise and fall" of 
intermediate-mass fragment production \cite{Bot95,Schue}, evolution of the
fragment mass and multiplicity distributions with excitation energy 
\cite{Dag, Schar}, fragment correlations revealing the critical behavior 
\cite{Dag, Schar}, confirmation of anomaly in the caloric curve \cite{Poch}, 
isoscaling \cite{Bot02}. Recent experiments \cite{GANIL}
directly confirm the basic assumption of the SMM, namely, that the
primary fragments are hot, their internal excitation energy may
reach up to 3 MeV per nucleon. Therefore, properties of these hot
nuclei can be extracted from multifragmentation reactions and used
for the description of matter under stellar conditions. The first
steps in this direction were made in our papers
\cite{Botvina04,Botvina05}. A similar model was also used in ref.
\cite{Japan} where, however, only cold nuclei in long-lived states
were considered.

\section{Statistical description of supernova matter}

\subsection{General remarks}

In the supernova environment, as compared to the multifragmentation 
reactions, several new important ingredients should be taken
into consideration. First, the matter at stellar scales must be
electrically neutral, and therefore electrons should be included to
balance positive nuclear charge.
Second, energetic photons present in hot matter may change
nuclear composition via photonuclear reactions.
And third, the matter is irradiated by a strong neutrino wind from the
protoneutron star.

Below we consider macroscopic volumes of matter consisting of
various nuclear species $(A,Z)$, nucleons $(n=(1,0)$ and
$p=(1,1))$, electrons $(e^-)$ and  positrons $(e^+)$ under the
condition of electric neutrality. We expect that in this situation
an  equilibrium ensemble of various nuclear species will be generated like in
a liquid-gas coexistence region, as observed in the
multifragmentation reactions. Now our system is
characterized by the temperature $T$, baryon density $\rho_B$ and
electron fraction $Y_e$ (i.e. the ratio of the net electron
density to the baryon density). One may expect that 
the new nuclear effects come into force in this environment. For example, the
liquid-drop properties in hot nuclei may be different from those
observed in cold nuclei (see discussion e.g. in refs.
\cite{Bot02,Bot06,Bot08}).

\subsection{Equilibrium conditions}

Composition of stellar matter can safely be studied within the Grand
Canonical Ensemble dealing with chemical potentials of the constituents.
Generally, the chemical potential of a species $i$ with baryon number $B_i$, charge $Q_i$
and lepton number $L_i$, which participates in chemical equilibrium,
can be found from the general expression:
\begin{equation}
\mu_i=B_i\mu_B+Q_i\mu_Q+L_i\mu_L
\end{equation}
where $\mu_B$, $\mu_Q$ and $\mu_L$ are three independent chemical
potentials which are determined from the conservation of total baryon
number $B=\sum_iB_i$ electric charge $Q=\sum_iQ_i$ and lepton number
$L=\sum_iL_i$ of the system. This gives
\begin{equation}
 \begin{array}{ll}
\mu_{AZ}=A\mu_B+Z\mu_Q~,~\\
\mu_{e^-}=-\mu_{e^+}=-\mu_Q+\mu_L~,~\\
\mu_\nu=-\mu_{\tilde{\nu}}=\mu_L~.
 \end{array}
\end{equation}
These relations are also valid for nucleons, $\mu_n=\mu_B$ and
$\mu_p=\mu_B+\mu_Q$.
If $\nu$ and $\overline{\nu}$ escape freely from the system, the
lepton number conservation is irrelevant and $\mu_L=0$. In this
case two remaining chemical potentials are determined from the
conditions of baryon number conservation and electro-neutrality:
\begin{equation}
\rho_B=\frac{B}{V}=\sum_{AZ}A\rho_{AZ}~,~
\rho_Q=\frac{Q}{V}=\sum_{AZ}Z\rho_{AZ}-\rho_e=0~.\nonumber
\end{equation}
Here $\rho_{AZ}$ is the number density of nuclear species $(A,Z)$,
$\rho_e=\rho_{e^-}-\rho_{e^+}$ is the net electron density.
The pressure of the relativistic electron-positron gas can be
written as
\begin{equation}
P_e=\frac{\mu_e^4}{12\pi^2}\left[1+2\left(\frac{\pi
T}{\mu_e}\right)^2+\frac{7}{15}\left(\frac{\pi T}{\mu_e}\right)^4
-\frac{m_e^2}{\mu_e^2}\left(3+\left(\frac{\pi
T}{\mu_e}\right)^2\right) \right]~,
\end{equation}
where the first order correction due to the finite electron mass is
included. The net number density $\rho_e$ and entropy density
$s_e$ can be obtained now from standard thermodynamic relations as
$\rho_e=\partial P_e/\partial \mu_e$ and $s_e=\partial
P_e/\partial T$. Neutrinos are taken into account in the same way,
but as massless particles, and with the spin factor twice smaller
than the electron one. The photon pressure is $P_{\gamma}=(\pi^2/45)T^4$.

\subsection{Nuclear statistical ensemble}

For describing an ensemble of nuclear species in thermodynamical
equilibrium we use the Grand Canonical version of the SMM
\cite{SMM,Bot85}, properly modified for supernova conditions.
After integrating out translational degrees of freedom the density
of nuclear species with mass $A$ and charge $Z$ is calculated as
\begin{equation} \label{naz}
\rho_{Az}=\frac{N_{AZ}}{V}=g_{AZ}\frac{V_f}{V}\frac{A^{3/2}}{\lambda_T^3}
{\rm exp}\left[-\frac{1}{T}\left(F_{AZ}-\mu_{AZ}\right)\right],
\end{equation}
were $g_{AZ}$ is the g.-s. degeneracy factor of species $(A,Z)$,
$\lambda_T=\left(2\pi\hbar^2/m_NT\right)^{1/2}$ is the nucleon
thermal wavelength, $m_N \approx 939$ MeV is the average nucleon
mass. $V$ is the actual volume of the system and $V_f$ is so
called free volume, which accounts for the finite size of nuclear
species. We assume that all nuclei have normal nuclear density
$\rho_0$, so that the proper volume of a nucleus with mass $A$ is
$A/\rho_0$. At low densities the finite-size correction can be
taken into account within the excluded volume approximation $V_f/V
\approx \left(1-\rho_B/\rho_0\right)$.

The internal excitations of nuclear species $(A,Z)$ play an
important role in regulating their abundance.  Sometimes they are
included through the population of nuclear levels known for nearly
cold nuclei (see e.g. \cite{Japan}). However, in the supernova
environment not only the excited states but also the binding
energies of nuclei will be strongly affected by the surrounding
matter. By this reason, we find it more justified to use another
approach which can easily be generalized to include in-medium
modifications of nuclear properties. Namely, the internal free energy of species $(A,Z)$
with $A>4$ is parameterized in the spirit of the liquid drop model
\begin{equation}
F_{AZ}(T,\rho_e)=F_{AZ}^B+F_{AZ}^S+F_{AZ}^{\rm sym}+F_{AZ}^C~~,
\end{equation}
where the right hand side contains, respectively, the bulk,
the surface, the symmetry and the Coulomb terms. The first three terms
are written in the standard form \cite{SMM},
\begin{eqnarray}
F_{AZ}^B(T)=\left(-w_0-\frac{T^2}{\varepsilon_0}\right)A,~
F_{AZ}^S(T)=\beta_0\left(\frac{T_c^2-T^2}{T_c^2+T^2}\right)^{5/4}A^{2/3},~
F_{AZ}^{\rm sym}=\gamma \frac{(A-2Z)^2}{A}. \nonumber
\end{eqnarray}
Here $w_0=16$ MeV, $\varepsilon_0=16$ MeV, $\beta_0=18$ MeV,
$T_c=18$ MeV and $\gamma=25$ MeV are the model parameters which
are extracted from nuclear phenomenology and provide a good
description of multifragmentation data
\cite{SMM,Bot95,Dag,Schar,GANIL}. However, some parameters,
especially $\gamma$, can be different in hot neutron-rich nuclei,
and they need more precise determination in nuclear experiments
(see e. g. ref. \cite{LeFev}). In the Coulomb term we include the
modification due to the screening effect of electrons. By using
the Wigner-Seitz approximation it can be expressed as \cite{Lat}
\begin{equation}
F_{AZ}^C(\rho_e)=\frac{3}{5}c(\rho_e)\frac{(eZ)^2}{r_0A^{1/3}}~,~~
c(\rho_e)=\left[1-\frac{3}{2}\left(\frac{\rho_e}{\rho_{0p}}\right)^{1/3}
+\frac{1}{2}\left(\frac{\rho_e}{\rho_{0p}}\right)\right]~,\nonumber
\end{equation}
where $r_0=1.17$ fm and $\rho_{0p}=(Z/A)\rho_0$ is the proton
density inside the nuclei. The screening function $c(\rho_e)$ is 1
at $\rho_e=0$ and 0 at $\rho_0=\rho_{0p}$. We want to stress that
both the reduction of the surface energy due to the finite
temperature and the reduction of the Coulomb energy due to the
finite electron density favor the formation of heavy nuclei.
Nucleons and light clusters $(A \leq 4)$ are considered as
structureless particles characterized only by mass and proper
volume.

\begin{figure}
\vspace{-0.5cm}
\includegraphics[width=12cm]{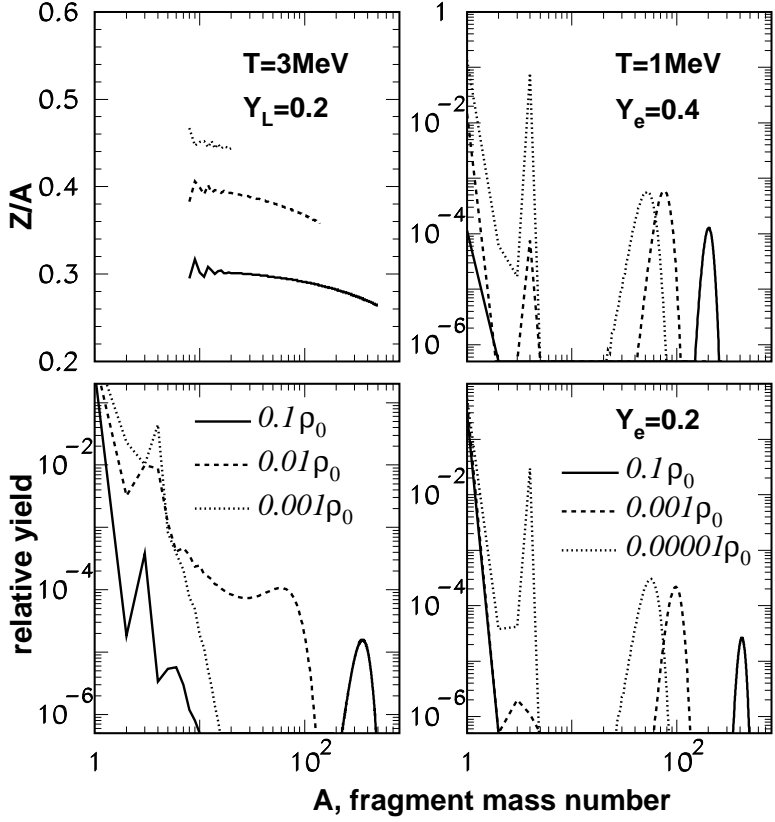}
\caption{\small{ Mean charge-to-mass ratios (left top panel), and
mass distributions of hot nuclei (other panels)
calculated with the SMM generalized for supernova conditions. Left
panels present calculations for temperature $T=3$ MeV and fixed lepton
(electrons+neutrinos) fraction $Y_L=$0.2 per nucleon. Right panels
are calculations for temperature $T=1$ MeV and fixed electron
fractions $Y_e=0.4$ (top) and 0.2 (bottom). Lines show the
fragment mass distributions at different baryon densities (in
units of the normal nuclear density $\rho_0$=0.15 fm$^{-3}$),
indicated in the figure. }}
\end{figure}

The pressure associated with nuclear species is calculated as for
the mixture of ideal gases,
\begin{equation} \label{pre}
P_{\rm nuc}=T\sum_{AZ}\rho_{AZ}\equiv
T\sum_{AZ}g_{AZ}\frac{V_f}{V}\frac{A^{3/2}}{\lambda_T^3} {\rm
exp}\left[-\frac{1}{T}\left(F_{AZ}-\mu_{AZ}\right)\right]~.
\end{equation}

As follows from eq. (\ref{naz}), the fate of heavy nuclei depends
sensitively on the relationship between $F_{AZ}$ and $\mu_{AZ}$.
In order to avoid an exponentially divergent contribution to the
baryon density, at least in the thermodynamic limit ($A
\rightarrow \infty$), inequality $F_{AZ}\goo \mu_{AZ}$ must hold.
The equality sign here corresponds to the situation when a big,
ultimately infinite,  nuclear fragment coexists with the gas of
smaller clusters \cite{Bug}. When $F_{AZ}>\mu_{AZ}$ only small
clusters with nearly exponentially falling mass spectrum are present.
However, there exist thermodynamic conditions corresponding to
$F_{AZ}\approx\mu_{AZ}$ when the mass distribution of nuclear
species is broadest. The advantage of our approach is that we
consider all the fragments present in this transition region,
contrary to the previous calculations \cite{Lamb,Lattimer}, which
consider only one ``average'' nucleus characterizing the liquid
phase.
\begin{figure}
\vspace{-1cm}
\includegraphics[width=12cm]{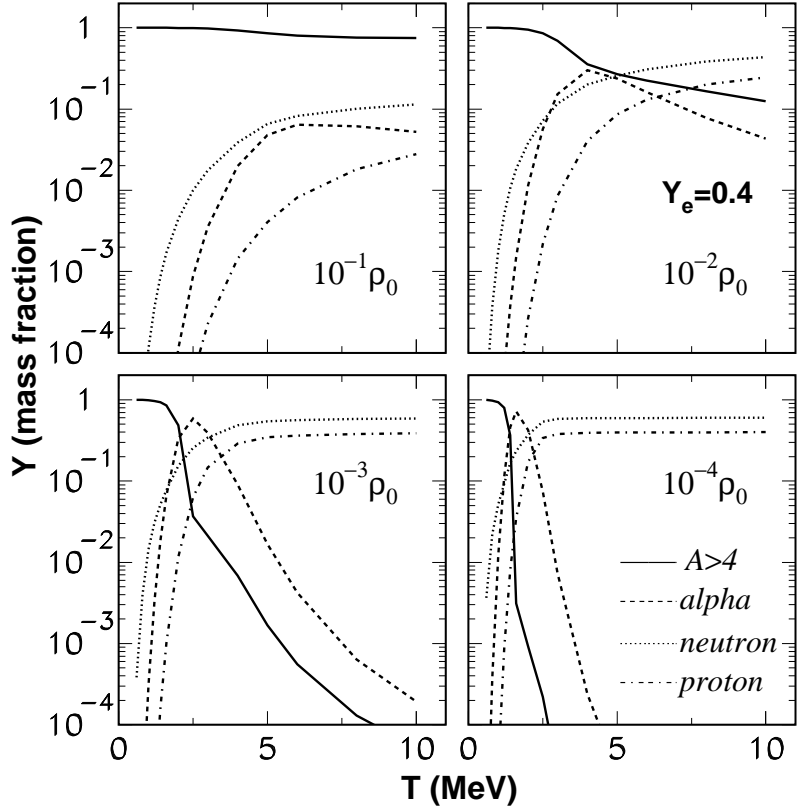}
\caption{\small{Mass fractions of different nuclear species as
functions of temperature for $Y_e=0.4$ calculated for different
baryon densities (indicated in the panels). Neutrons, protons,
$\alpha$-particles and heavier nuclei (A$>$4) are shown by dotted,
dash-dotted, dashed and solid lines, respectively.}}
\end{figure}

\section{Numerical results}

\subsection{Nuclear composition}
In numerical calculations we first fix temperature $T$, baryon
density $\rho_B$ and electron fraction $Y_e$. Then we consider a
box containing the baryon number $B=$1000 and proton number
$Z=Y_e\cdot B$. The box volume is fixed by the average baryon
density, $V=B/\rho_B$. We use an iterative procedure to find
chemical potentials $\mu_B$ and $\mu_Q$. Finally, relative yields
of all nuclei with 1$\leq A \leq$1000 and 0$\leq Z \leq A$ are
calculated from eq. (\ref{naz}). Nuclei with larger masses
($A>$1000) can be produced only at relatively high densities,
$\rho_B>0.1\rho_0$, which are relevant for the regions deep inside
the protoneutron star, and which are not considered here.
\begin{figure}
\vspace{-1cm}
\includegraphics[width=13cm]{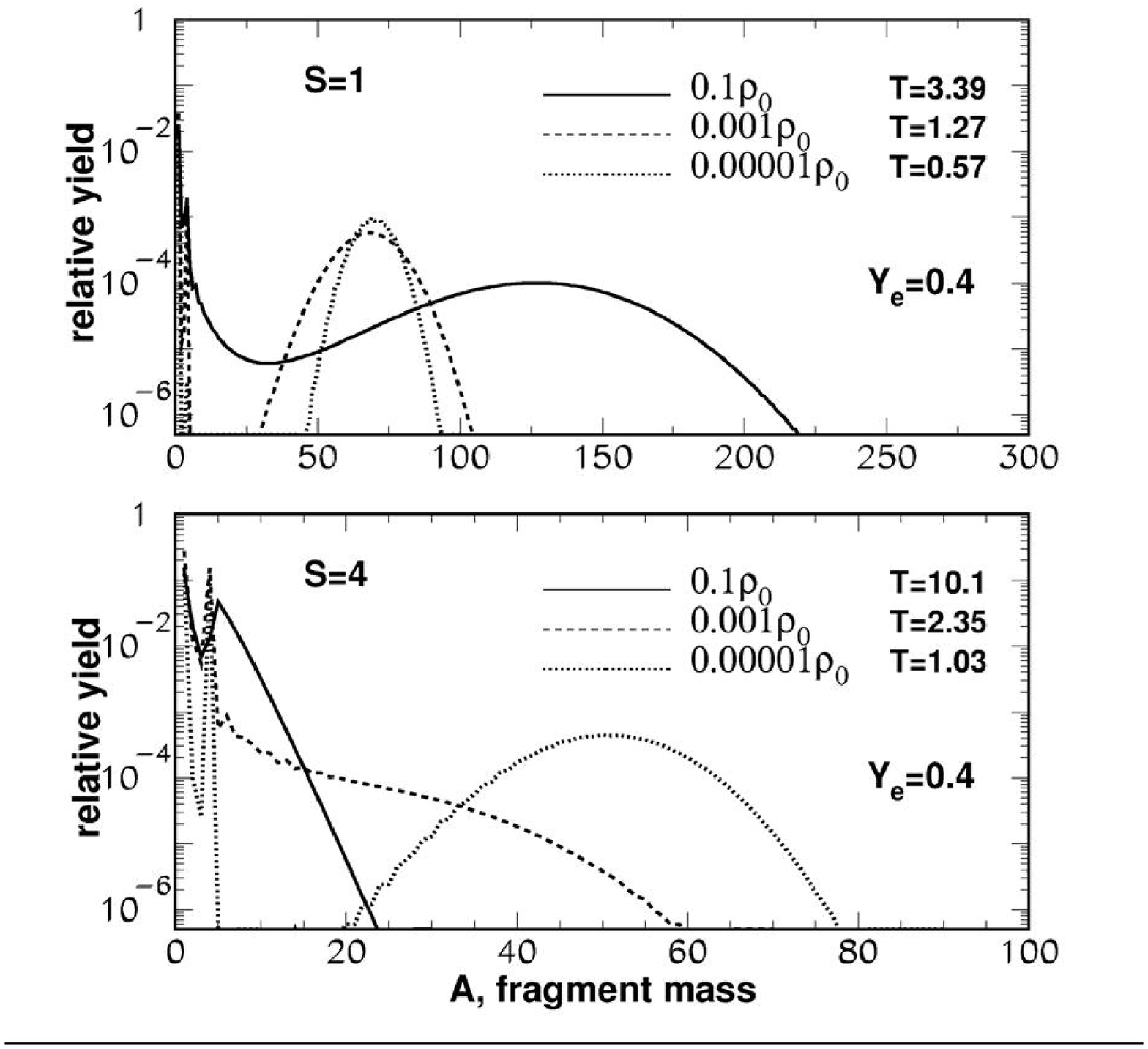}
\caption{\small{Mass distributions of nuclear species along two
isentropes with entropy per baryon equal 1.0 (upper panel) and 4.0
(lower panel). The corresponding temperatures and densities are
indicated in the figure.} }
\end{figure}
First we consider the case when lepton fraction is fixed as
expected inside a neutrinosphere. Figure~2 (left panels) shows the
results for lepton fraction $Y_L$=0.2 and typical temperature
$T=3$ MeV.  Mass distributions are shown in the lower left panel.
One can see that the islands of heavy nuclei, $200<A<400$, appear
at relatively high baryon density, $\rho_B=0.1\rho_0$,
corresponding to the vicinity of a protoneutron star. These nuclei
are very neutron-rich, $Z/A\approx 0.27$. The $Z/A$ ratios are
decreasing with $A$ less rapidly than in the nuclear
multifragmentation case \cite{Bot01}. This can be explained by the
screening effect of electrons. The width of the charge
distribution at given $A$ is determined by $T$ and $\gamma$:
$\sigma_Z \approx \sqrt{AT/8\gamma}$ \cite{Bot85,Bot01}. At lower
density, $\rho_B=0.01\rho_0$, the mass distribution is rather flat
up to $A\approx 80$ and then decreases rapidly for larger $A$. For
$\rho_B=10^{-3}\rho_0$ only light clusters are present and the
mass distribution drops exponentially.

Let us consider now the situation more appropriate for a hot bubble at
early times of a supernova explosion, when the electron fraction of matter
did not change significantly by the electron capture reactions.
In this case the electron fraction is fixed to the initial value, 
and the electron and proton chemical potentials are determined
independently, without using the equilibrium relation
$\mu_e=-\mu_Q$. Corresponding results for $Y_e=0.4$ and $Y_e=0.2$
at $T=1$ MeV and several baryon densities are presented in Fig.~2
(left top and bottom panels).

\begin{figure}
\vspace{-1cm}
\includegraphics[width=12cm]{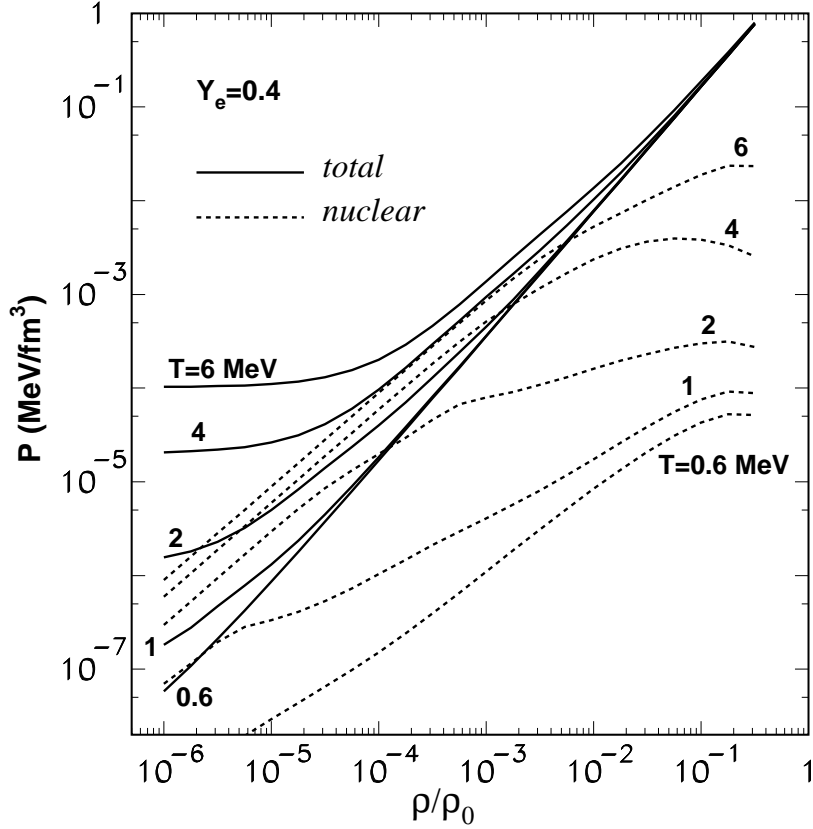}
\caption{\small{Pressure isotherms as functions of relative baryon density for $Y_e$=0.4. Solid lines show the total pressure including the electron, photon and nuclear contributions. Dotted lines show only the nuclear contribution. Results are presented for temperatures 6, 4, 2, 1 and 0.6 MeV (from top to bottom), as indicated at the corresponding lines.}}
\end{figure}
One can see that heavy and even superheavy nuclei, $50<A<400$, can
be formed in this case too. They exist in a very broad range of
densities, $0.1\rho_0>\rho_B>10^{-5}\rho_0$. At given density the
mass distribution of heavy nuclei has a Gaussian shape. In the $Y_e=0.4$ 
case the most probable nuclei, corresponding to the maxima of
distributions, have $Z/A$ ratios  0.400, 0.406, and 0.439, for
densities $0.1\rho_0$, $10^{-3}\rho_0$, and $10^{-5}\rho_0$,
respectively. The Gaussian mass distributions may in some cases
justify earlier calculations \cite{Lamb,Lattimer}, when only one kind of 
nuclear species was considered at each density. As seen from
the bottom panel, changing the electron fraction from 0.4 to 0.2
leads to a significant increase of nuclear masses. Also, the
nuclei become more neutron rich: the corresponding $Z/A$ ratios
are 0.280, 0.359, and 0.420. Our calculations show that even
larger effect can be caused by the reduction of the symmetry
energy of hot fragments (see ref. \cite{Bot01}). 

Figure 3 displays the mass fractions of different nuclear species
as functions of temperature for several baryon densities and fixed
$Y_e=0.4$. One can see several interesting trends. First, nuclei
with $A>4$ survive at high temperatures only if the baryon density
is large enough, $\rho_B>10^{-2}\rho_0$, At lower densities they
are destroyed by hard photons already at $T>2$ MeV. On the other
hand, the neutron and proton fractions increase gradually and
dominate at $\rho_B\leq 10^{-2}\rho_0$. A significant change in
the trend is observed at $T>3$ MeV which can be related to the
liquid-gas transition in such a matter. It is interesting to note
that $\alpha$-particles may exist abundantly only in a narrow
range of temperatures, $2<T<4$ MeV (see two lower panels).

\subsection{Isentropic trajectories}
let us consider now how the composition of matter changes along
the isentropic trajectories. Fig.~4 displays the mass
distributions of nuclear species along two isentropes, $S/B=1.0$
and 4.0. One can clearly see the different trends in these
two cases.  In the first case the widest distribution corresponds
to the highest temperature and density state, $T=3.39$ MeV,
$\rho_B=10^{-1}\rho_0$. The mass distribution extends up to about
$A=230$ in this case. At lower densities the mass distributions
are peaked at $A\approx 70$. However, at $S/B=4.0$ the nuclei are
generally much lighter, and the widest distribution corresponds to
the lowest density state, $\rho_B=10^{-3}\rho_0$, $T=1.03$ MeV. It
is remarkable and somewhat unexpected that relatively heavy nuclei
with $20<A<80$ can survive at such a high specific entropy.

One should bear in mind that the mass distributions which are
presented here correspond to hot primary nuclei. After ejection
these nuclei will undergo de-excitation. At typical temperatures
considered here ($T\loo 3$ MeV) the internal excitation energies
are relatively low, less than 1.0 MeV/nucleon. As well known from
calculations \cite{SMM} and nuclear experiments
\cite{Dag,Schar,GANIL}, de-excitation of nuclei with $A\leq$ 200
will go mainly by means of the nucleon emission. Then the
resulting distributions of cold nuclei are not very different from
the primary ones, they are shifted to lower masses by several
units. One should expect that shell effects (which, however, may
be modified by surrounding electrons) will play an important role
at the de-excitation stage leading to the fine structure of the
mass distribution. We believe that after the de-excitation of hot
nuclei, corresponding to the time when the ejected matter reaches
very low densities, the r-process may be responsible for the final
redistribution of the element abundances, leading to the
pronounced peaks around $A\approx$80, 130 and 200 \cite{Qui}.

As well known, nuclear
composition is extremely important for the physics of supernova
explosions. For example, the electron capture on nuclei plays an
important role in supernova dynamics \cite{Hix}. But the electron
capture rates are sensitive to the nuclear composition and details
of nuclear structure (see e.g. \cite{LMP}). The neutrino-induced
reactions are very sensitive to the nuclear structure effects and
properties of weak interactions in nuclei (see e.g. \cite{Hor}).
It is also important that the presence of nuclei favors the
explosion via the energy balance in the bubble \cite{Bethe}, since
more energy can be used for the explosion. All these
considerations show importance of the nuclear physics input in
supernova phenomenon.

\subsection{Equation of state}

Finally, we present results concering thermodynamical properties of supernova matter. 
Figure~6 shows the isothermic equation of state on the pressure---density plane. One can clearly see that the pressure is dominated by the relativistic electrons at high baryon densities and by thermal photons at low baryon densities. The nuclear contribution is is always small compared to these two contributions. 

On the other hand, the nuclear pressure shows the tendency to saturation at higher densities. This is consequence of the liquid-gas phase transition in nuclear subsystem, which in thermodynamic limit will manifest itself by a constant pressure in the coexistence region. This behavior will significantly influence the thermodynamical properties of matter, in particular, its heat capacity \cite{Bug}.  

\section{Conclusions}
\begin{itemize}

\item The statistical equilibrium approach, which was successfully used for describing  multifragmentation reactions, can be applied also for calculating the equation of state and nuclear composition of supernova matter. 

\item Survival of (hot) heavy nuclei may significantly influence the explosion dynamics through both the energy balance and modification of the weak reaction rates.

\item Statistical mechanism may provide "seed" nuclei for further nuclear transformations in r-, rp, and s- processes.

\item Due to the screening effect of electrons, the alpha-decay and spontaneous fission may be suppressed in supernova environments, that opens the pathway to the production of heavy and superheavy elements. 

\end{itemize}

I am grateful to A.S. Botvina with whom most of the presented results were obtained.
This work was supported in part by the DFG grant 436RUS 113/711/0-2 (Germany), and grants RFFR-05-02-04013 and NS-8756.2006.2 (Russia).


\end{document}